\documentclass[10pt,twocolumn,letterpaper]{article}

\usepackage{cvpr}              

\usepackage{graphicx}
\usepackage{amsmath}
\usepackage{amssymb}
\usepackage{booktabs}

\usepackage{algorithm}
\usepackage{algorithmic}

\usepackage[pagebackref,breaklinks,colorlinks]{hyperref}

\usepackage[capitalize]{cleveref}
\crefname{section}{Sec.}{Secs.}
\Crefname{section}{Section}{Sections}
\Crefname{table}{Table}{Tables}
\crefname{table}{Tab.}{Tabs.}

\usepackage{multirow}

\def\cvprPaperID{7682} 
\def\confName{CVPR}
\def\confYear{2023}

\begin{document}

\title{Adversarial Color Projection: A Projector-based Physical Attack to DNNs}




\author{Chengyin Hu\\
University of \\Electronic Science and\\ Technology of China\\
Chengdu, China\\
{\tt\small cyhuuestc@gmail.com}
\and
Weiwen Shi\\
University of \\Electronic Science and\\ Technology of China\\
Chengdu, China\\
{\tt\small Weiwen\_shi@foxmail.com}
\and
Ling Tian \thanks{Corresponding author}\\
University of \\Electronic Science and\\ Technology of China\\
Chengdu, China\\
{\tt\small lingtian@uestc.edu.cn}
}

\maketitle

\begin{abstract}
Recent research has demonstrated that deep neural networks (DNNs) are vulnerable to adversarial perturbations. Therefore, it is imperative to evaluate the resilience of advanced DNNs to adversarial attacks. However, traditional methods that use stickers as physical perturbations to deceive classifiers face challenges in achieving stealthiness and are susceptible to printing loss. Recently, advancements in physical attacks have utilized light beams, such as lasers, to perform attacks, where the optical patterns generated are artificial rather than natural. In this work, we propose a black-box projector-based physical attack, referred to as adversarial color projection (\textbf{AdvCP}), which manipulates the physical parameters of color projection to perform an adversarial attack. We evaluate our approach on three crucial criteria: effectiveness, stealthiness, and robustness. In the digital environment, we achieve an attack success rate of 97.60\% on a subset of ImageNet, while in the physical environment, we attain an attack success rate of 100\% in the indoor test and 82.14\% in the outdoor test. The adversarial samples generated by AdvCP are compared with baseline samples to demonstrate the stealthiness of our approach. When attacking advanced DNNs, experimental results show that our method can achieve more than 85\% attack success rate in all cases, which verifies the robustness of AdvCP. Finally, we consider the potential threats posed by AdvCP to future vision-based systems and applications and suggest some ideas for light-based physical attacks.

\end{abstract}


\begin{figure}
\centering
\includegraphics[width=1\columnwidth]{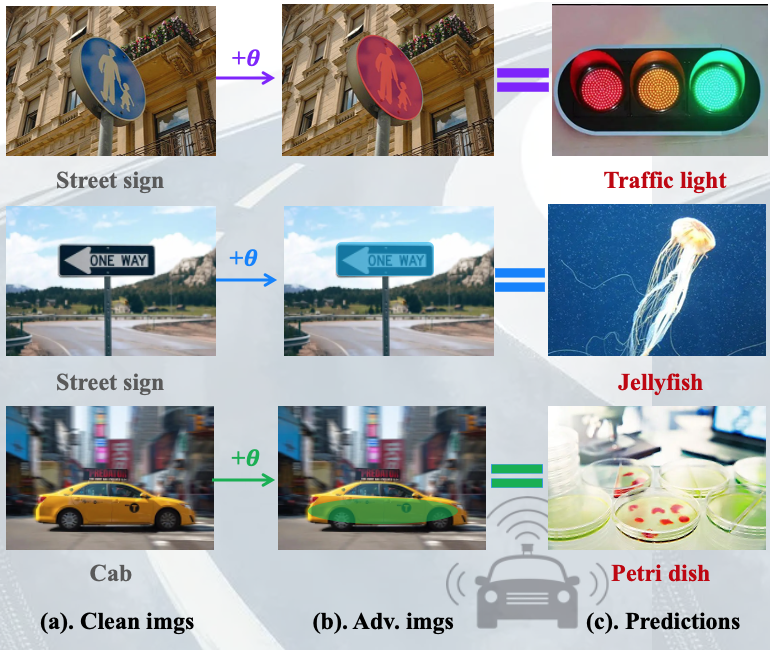} 
\caption{An example. When a camera installed in a self-driving car captures target objects that have been subjected to a carefully designed color projection attack, the object recognition system may fail to correctly identify the sign as a "Street sign" or the other objects, such as "Cab".}.
\label{figure1}
\end{figure}

\section{Introduction}
\label{sec1}
In recent years, deep neural networks (DNNs) have achieved significant progress in numerous tasks, such as image classification, object detection, and domain segmentation. Concurrently, vision-based applications are becoming increasingly popular in people's daily lives, including UAV and autonomous driving. Consequently, the security and reliability of vision-based systems have become a key focus for many scholars \cite{ref59,ref61}. While most scholars are primarily focused on adversarial attacks in the digital environment \cite{ref82,ref83}, where subtle perturbations are intentionally added to input images and fed to the target model, and these perturbations are difficult to detect by human observers, some scholars are gradually shifting their attention to physical attacks \cite{ref86,ref87}. Unlike digital attacks, physical attacks involve capturing images via cameras and feeding them to the target model. Physical perturbations are usually designed to be more significant to ensure they are captured by the camera. However, they also need to be carefully designed to be undetectable to human observers, making physical attacks a compromise between stealthiness and robustness. 

Natural phenomena can serve as physical perturbations, for example, a bright light beam can cause a self-driving car to crash. However, what if we utilize color beams as physical perturbations? Specifically, we can use color beams as physical perturbations to exploit the transience of light-speed attacks and perform physical attacks on advanced DNNs. To simulate various color beams, we can use a projector as a light device and project color beams as physical perturbations. As shown in Figure \ref{figure1}, a carefully designed color projection can be projected onto a street sign by the attacker, causing the autonomous vehicle to fail to classify it accurately.

\begin{figure}
\centering
\includegraphics[width=1\columnwidth]{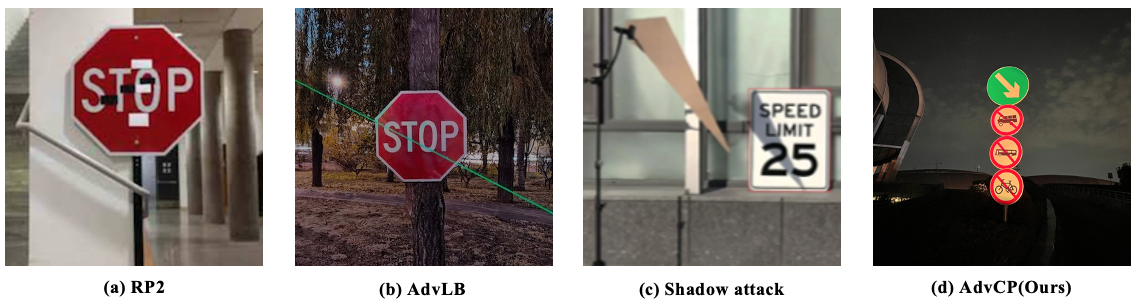} 
\caption{Visual comparison.}.
\label{figure2}
\end{figure}

Currently, most physical attacks rely on stickers as physical perturbations, which are attached to target objects to execute attacks \cite{ref24,ref26}. Such methods can achieve effective physical attacks without altering the semantic information of the target objects. However, because the sticker is applied to the surface of the object, it is difficult to achieve stealthiness with sticker-based attacks. Furthermore, some researchers have explored physical attacks using light beams as physical perturbations \cite{ref33,ref34,ref35,ref51}, which exploit the transience of light to achieve stealthy physical attacks at the cost of reduced robustness. Another approach is camera-based physical attacks \cite{ref38}, in which some tiny translucent patches are placed on the camera of a mobile phone to perform physical attacks. However, adding too many patches can result in large experimental errors.

In this study, we present a novel physical attack approach called adversarial color projection, which utilizes a projector to project a carefully designed color beam onto target objects instead of physically attaching stickers to them, making our approach more versatile than most existing methods. In comparison to current light-based physical attacks, our approach is more natural and provides improved stealthiness. To illustrate this, we provide a visual comparison in Figure \ref{figure2}, which highlights that the physical samples generated by AdvCP are significantly more stealthy than those created by other techniques such as RP2 \cite{ref24}, AdvLB \cite{ref35}, and shadow attack \cite{ref37}.

Our proposed method offers an easy and effective way to deploy physical attacks by formalizing the physical parameters of color projection and utilizing particle swarm optimization algorithm \cite{ref50} to determine the most aggressive parameters. The resulting color projection is then used to generate physical samples on the target objects by a projector. To summarize, our main contributions are as follows:

\begin{itemize}
\item In this paper, we introduce a novel physical attack method called AdvCP that leverages the unique properties of light to perform a black-box attack. Compared to existing physical attacks, AdvCP offers greater flexibility and convenience, and therefore poses a more significant threat.  (See Section \ref{sec1}.)

\item We introduce and analyze existing methods (See Section \ref{sec2}), design rigorous methods, and perform comprehensive experiments to verify the effectiveness, stealthiness and robustness of AdvCP (See Section \ref{sec3}, \ref{sec4}).Given the successful testing of AdvCP in real-world scenarios, it has the potential to serve as a valuable tool for investigating the threat of light-based attacks in practical settings.

\item We conduct a comprehensive analysis of AdvCP, including ablation study and defense strategies (See Section \ref{sec5}). Additionally, we propose some new ideas for light-based physical attacks(See Section \ref{sec6}).

\end{itemize}

\section{Related work}
\label{sec2}
\subsection{Digital attacks}

Adversarial attacks were initially proposed by Szegedy et al. \cite{ref1}, who demonstrated that advanced DNNs could be easily manipulated by small perturbations, paving the way for successful adversarial attacks \cite{ref84,ref85}.

Most digital attacks limit the size of the adversarial perturbations to ensure that they are imperceptible to human observers. The ${l}_{2}$ and ${l}_{\infty}$ norms are the most commonly used methods for measuring perturbations \cite{ref88,ref89}. Additionally, some researchers modify other attributes of digital images, such as color \cite{ref8,ref9}, texture, and camouflage \cite{ref90,ref91,ref13}, to generate perturbations that are slightly perceptible to human observers. Other researchers modify the physical parameters of digital images \cite{ref14,ref15} and retain only the key components of the images to generate adversarial samples. Several works \cite{ref79, ref80} have proposed the raindrop attack, which utilizes simulated raindrops as perturbations to evaluate its effectiveness in attacking deep neural networks and to develop defense mechanisms to enhance their robustness. In general, the assumption of digital attacks is that an attacker can modify the input image, but this is not practical in a physical scenario.

\subsection{Physical attacks}

The concept of physical attacks was first introduced by Kurakin et al. in \cite{ref22}, which subsequently paved the way for many successful physical attacks \cite{ref92,ref28}.

\textbf{Traditional street sign attacks.} Ivan Evtimov et al. \cite{ref24} proposed a classical physical attack called RP2, which utilizes stickers as perturbations to launch attacks at various distances and angles against advanced DNNs. However, RP2 is susceptible to environmental interference at large distances and angles. Eykholt et al. \cite{ref26} improved RP2 by implementing a disappear attack, generating robust and transferable adversarial samples to deceive advanced DNNs. However, the perturbations cover a large area, making them too conspicuous. Chen et al. \cite{ref23} proposed ShapeShifter, which utilizes "Expectation over Transformation" to generate adversarial samples. Experimental results showed that the generated stop sign consistently deceived advanced DNNs at various distances and angles. Huang et al. \cite{ref27} further improved ShapeShifter by incorporating Gaussian white noise to the optimization function, resulting in a more comprehensive attack. However, both ShapeShifter and its improved version suffer from a defect wherein the perturbations cover almost the entire road sign, failing to achieve stealthiness. Duan et al. \cite{ref25} proposed AdvCam, which employs style transfer techniques to generate adversarial samples and disguise the perturbations as a style that is perceived as reasonable by human observers. Although this method is more stealthy than the above methods, it still requires manual selection of the target area. In summary, the above methods necessitate manual modification of the target objects, and the perturbations covering almost the entire road sign lead to a lack of stealthiness.

\textbf{Physical attacks against face recognition system.} Nguyen et al. \cite{ref33} proposed a novel approach to attacking face recognition systems using a projector to generate adversarial samples. They verified the effectiveness of this approach under both white-box and black-box settings. However, this method requires a complex deployment mode. Additionally, other studies \cite{ref32,ref34} have utilized visible and invisible light to perform stealthy attacks on face recognition systems by projecting optimized light onto the target. These methods not only achieve effective attacks on face recognition systems, but also improve the stealthiness of physical attacks. 

\textbf{Laser-based attack.} Duan et al. \cite{ref35} proposed a novel physical attack technique called AdvLB that utilizes laser beams as perturbations and manipulates their physical parameters to execute attacks. Compared to traditional street sign attacks, AdvLB offers greater flexibility. Additionally, due to the nature of light, AdvLB can achieve better stealthiness. However, it is important to note that AdvLB is susceptible to displacement errors in physical attack scenarios.

\textbf{Projector-based attack.} Gnanasambandam et al. \cite{ref36} proposed a technique called OPAD that amplifies subtle digital perturbations and projects them onto the target object to generate physical adversarial samples. However, the irregular projection patterns produced by this method may raise suspicion among human observers. 

\textbf{Shadow-based attack.} Zhong et al. \cite{ref37} conducted a study on shadow-based physical attacks, in which they cast meticulously crafted shadows on the target object to generate adversarial samples, thereby achieving a natural black-box attack. Due to the ubiquity of shadows, this method offers enhanced stealthiness. However, the deployment of shadow-based attacks necessitates the placement of the cardboard close to the target object, which makes it susceptible to human observer suspicion.

\textbf{Camera-based attack.} Li et al. \cite{ref38} investigated camera-based attacks, in which they applied precisely designed stickers on the camera lens to generate inconspicuous adversarial samples for targeted attacks on advanced DNNs. This method circumvents the need to physically modify the target by manipulating the camera itself. However, placing multiple patches on the camera lens may lead to significant errors in the attack.

\textbf{Raindrop-based attack.} Guesmi et al. \cite{ref81} introduced AdvRain, a physical attack that employs simulated raindrops to generate adversarial samples. The authors demonstrated the efficacy of AdvRain by conducting experiments on two popular DNN models, VGG19 and ResNet34, and obtained an average model accuracy reduction of 45\% and 40\%, respectively, using only 20 raindrops. AdvRain is notable for its stealthiness in physical attacks, although further investigations are necessary to evaluate its robustness.

\begin{figure*}[t]
\centering
\includegraphics[width=1\linewidth]{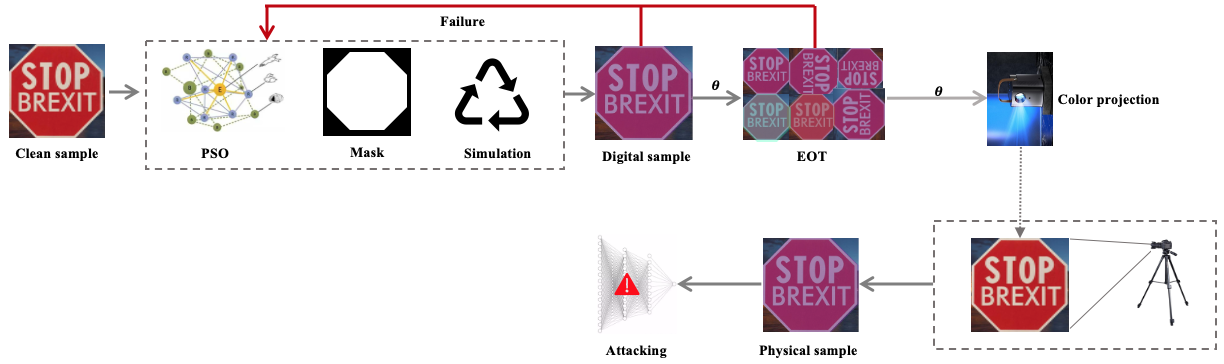} 
\caption{The attacker applies Particle Swarm Optimization (PSO) to optimize the simulation samples and uses Expectation Over Transformation (EOT) to transition the digital samples to physical samples.}.
\label{figure3}
\end{figure*}

\section{Approach}
\label{sec3}

\subsection{Adversarial sample}
Let $X$ be an input image with ground truth label $Y$, and $f$ be a deep neural network classifier. We denote $f(X)$ as the predicted label for $X$, and ${f}_{Y}(X)$ as the confidence score associated with class $Y$. An adversarial sample ${X}_{adv}$ satisfies two conditions: (1) $f({X}_{adv}) \neq f(X) = Y$, indicating that ${X}_{adv}$ fools the DNN classifier $f$; and (2) $\parallel{X}_{adv} - X\parallel < \epsilon$, ensuring that the perturbations in ${X}_{adv}$ are small enough to be imperceptible to human observers. To generate an adversarial sample, an attacker may use PSO to optimize the perturbations and EOT to transition the digital samples to physical samples. Finally, a color projection is used to project the adversarial color beams onto the target object, further deceiving the classifier.

In this study, we employ particle swarm optimization \cite{ref50} to optimize the physical parameters of the color projection. We then project the optimized color projection onto the target object using a projector in real-world scenarios to generate physical samples. Figure \ref{figure3} depicts our approach.

\subsection{Color projection definition}

In this paper, we define color projection using three physical parameters: location ${\mathcal{P}}_{l}$, color $\mathcal{C}(r, g, b)$, intensity $\mathcal{I}$. Each parameter is described as follows:

\textbf{Location ${\mathcal{P}}_{l}$:} For the location of color projection, we propose to use a polygon ${\mathcal{P}}_{l}$ with a set of vertices $l=\{({m}_{1},{n}_{1}),({m}_{2},{n}_{2}),...,({m}_{k},{n}_{k})\}$ to simulate the projection region, where $\mathcal{M}$ is the mask that locates the target object, and the color projection region is expressed as $\mathcal{M} \cap {\mathcal{P}}_{l}$. Although any polygon can be used for ${\mathcal{P}}_{l}$, we use full-coverage polygons, such as octagons, to ensure stealthiness in our physical attack. Using different polygons may increase the attack success rate, but it also makes the color projection more conspicuous. Experimental tests in Table \ref{Table 4} demonstrate that quadrangular color projection is sufficient to generate successful adversarial samples. Therefore, we use quadrangular color projections in simulation experiments and full-coverage color projections in physical environments for the attack.

\textbf{Color $\mathcal{C}(r, g, b)$:}  In our approach, $\mathcal{C}(r, g, b)$ refers to the color of the color projection. Specifically, the red, green, and blue channels of the color projection are denoted as $r$, $g$, and $b$, respectively.

\textbf{Intensity $\mathcal{I}$:} $\mathcal{I}$ indicates the intensity of color projection. 

The physical parameters ${\mathcal{P}}_{l}$, $\mathcal{C}(r, g, b)$, and $\mathcal{I}$ are combined to form a color projection's physical parameter $\theta(\mathcal{C}, {\mathcal{P}}_{l}, \mathcal{I})$. To synthesize an adversarial sample, we define a function $S(X; \theta(\mathcal{C}, {\mathcal{P}}_{l}, \mathcal{I}), \mathcal{M})$ that applies the color projection to the input image. The range of the physical parameters $\mathcal{C}$, ${\mathcal{P}}_{l}$, and $\mathcal{I}$ is limited by the restriction vectors ${\vartheta}_{min}$ and ${\vartheta}_{max}$, which can be adjusted. Thus, the adversarial sample can be expressed as:

\begin{equation}
    \label{Formula 1}
    {X}_{adv} = S(X; \theta(\mathcal{C}, {\mathcal{P}}_{l}, \mathcal{I}), \mathcal{M}) 
\end{equation}

$$s.t. \quad \theta(\mathcal{C}, {\mathcal{P}}_{l}, \mathcal{I}) \in ({\vartheta}_{min},{\vartheta}_{max})$$

In which, ${X}_{adv}$ refers to the adversarial samples obtained by applying the simple linear fusion method $S$ to the clean sample $X$ and the color projection $\theta$.

\textbf{Expectation Over Transformation (EOT).} EOT \cite{ref31} is an effective technique for dealing with the transition from digital to physical domains. In this work, we define a transformation $\mathcal{T}$ to represent the domain transition, which is a random combination of digital image processing techniques, including brightness adaptation, position offset, color variation, etc. By using EOT, the physical sample can be expressed as a function of the digital sample and the transformation:

\begin{equation}
    \label{Formula 2}
    {X}_{phy} = \mathcal{T}({X}_{adv}, \theta(\mathcal{C}, {\mathcal{P}}_{l}, \mathcal{I})) 
\end{equation}

$$s.t. \quad \theta(\mathcal{C}, {\mathcal{P}}_{l}, \mathcal{I}) \in ({\vartheta}_{min},{\vartheta}_{max})$$

The aim of AdvCP is to search for the physical parameters $\theta(\mathcal{C}, {\mathcal{P}}_{l}, \mathcal{I})$ of the color projection that can generate an adversarial sample ${X}_{adv}$, fooling the classifier $f$. In practical scenarios, attackers may not have access to the model, but can obtain the confidence score ${f}_{Y}(X)$ of the given input image $X$ on the ground truth label $Y$. In our proposed method, we leverage the confidence score as the adversarial loss, and formulate the objective as the minimization of the confidence score with respect to the ground truth label $Y$, which can be mathematically expressed as:

\begin{equation}
    \label{Formula 3}
    \mathop{\arg\min}_{\theta}{\mathbb{E}}_{t \sim \mathcal{T}}[{f}_{Y}(t({X}_{adv}, \theta(\mathcal{C}, {\mathcal{P}}_{l}, \mathcal{I})))]
\end{equation}

$$ s.t. \quad f({X}_{adv}) \neq Y$$

\subsection{Color projection adversarial attack}

Particle Swarm Optimization (PSO) \cite{ref50} is an algorithm that simulates the foraging behavior of birds. In PSO, a population of random particles is initialized, and the optimal solution is found through iterative updates. In each iteration, particles search for the optimal solution by tracking two extreme values: the best solution ${\theta}_{i,best}$ found by the individual particle and the best solution ${\theta}_{best}$ found by the entire population. When a particle satisfies the objective function, the solution to the problem is output. Notably, in this work, we do not require the gradient information of the target model, only the prediction label and the confidence score of the model are needed. The effectiveness of using PSO to optimize AdvCP can be attributed to the following factors:

(1) PSO is a versatile algorithm that does not rely on prior knowledge of the target problem. The effectiveness of using PSO to optimize AdvCP lies in its ability to optimize without using gradient information of the model. In particular, we define the fitness value as ${f}_{Y}(X)$ and the termination condition as $f({X}_{adv}) \neq Y$. This approach allows for a more flexible and adaptable optimization process, making PSO a valuable tool for AdvCP.

(2) PSO is a powerful algorithm that can search for optimal solutions across a wide range, making it particularly suitable for global optimization. In this work, the physical parameters of color projection $\theta$ have a vast solution space, and PSO's ability to explore the search space globally is leveraged to find the optimal solution for AdvCP.

(3) PSO is known for its ability to efficiently explore a large solution space and find the global optimum while avoiding local optima. This is due to the leap of PSO, which allows particles to move quickly towards better solutions. In our proposed method, we leverage the flexibility of PSO to select suitable parameters and obtain the global optimal solution for the defined parameters, ultimately leading to improved performance in the optimization of AdvCP.

Next we describe the process of optimizing AdvCP using PSO.

\textbf{Initialization.} We initialize the swarm and the velocity:

\begin{equation}
    \label{Formula 7}
    POP=[{\theta}_{1},{\theta}_{2},...,{\theta}_{I}]
\end{equation}

\begin{equation}
    \label{Formula 8}
    V=[{v}_{1},{v}_{2},...,{v}_{I}]
\end{equation}
where $I$ denotes the population size, ${\theta}_{i}$ ($i=1,2,...,I$) denotes a candidate solution in $POP$, ${v}_{i}$  denotes the moving direction of particle ${\theta}_{i}$.

\textbf{Obtaining the individual optimal and population optimal solutions.} In this stage, the individual optimal solution and the population optimal solution from the initial population to the current population are obtained.

\begin{equation}
\begin{split}
    \label{Formula 9}{\theta}_{i,best}^{j}=\mathop{\arg\min}_{{\theta}_{i,best}^{u}}{\mathbb{E}}_{t \sim \mathcal{T}}[{f}_{Y}(t({X}_{adv}, {\theta}_{i,best}^{u}))] \\  u=1,2...,j
\end{split}
\end{equation}

\begin{algorithm}[t]
	\renewcommand{\algorithmicrequire}{\textbf{Input:}}
	\renewcommand{\algorithmicensure}{\textbf{Output:}}
	\caption{Pseudocode of AdvCP}
	\label{algorithm1}
	\begin{algorithmic}[1]
	
		\REQUIRE Input $X$, Classifier $f$, Ground truth label $Y$, Max step ${t}_{max}$, $\omega$, ${c}_{1}$, ${c}_{2}$;
		\ENSURE A vector of parameters $\theta$;

        \STATE Initialization: $POP=[{\theta}_{1},{\theta}_{2},...,{\theta}_{I}]$;\\ $V=[{v}_{1},{v}_{2},...,{v}_{I}]$;\\

        \FOR{$j$ $\leftarrow$ 0 to ${t}_{max}$}
        \FOR{each ${\theta}_{i}^{j}$ in $POP$}
            \STATE ${X}_{i}^{j}=S(X,{\theta}_{i}^{j}(\mathcal{C}, {\mathcal{P}}_{l}, \mathcal{I}), \mathcal{M})$;
            \STATE ${f}_{Y}({X}_{i}^{j}) \leftarrow f({X}_{i}^{j})$;
            \STATE ${\theta}_{i,best}^{j}=\mathop{\arg\min}_{{\theta}_{i,best}^{u}}{\mathbb{E}}_{t \sim \mathcal{T}}[{f}_{Y}(t({X}_{adv}, {\theta}_{i,best}^{u}))]$ \\ $u=1,2...,j$;
            \IF{$f({X}_{i}^{j}) \neq Y$}
                \STATE Output $\theta = {\theta}_{i}^{j}(\mathcal{C}, {\mathcal{P}}_{l}, \mathcal{I})$;
                \STATE break;
            \ENDIF
        \ENDFOR
        \STATE ${\theta}_{best}^{j}=\mathop{\arg\min}_{{\theta}_{i,best}^{u}}{\mathbb{E}}_{t \sim \mathcal{T}}[{f}_{Y}(t({X}_{adv}, {\theta}_{i,best}^{u}))]$ \\  $i=1,2,...,I \quad u=1,2,...,j$;
        
        \FOR{each ${\theta}_{i}^{j}$ in $POP$}
        \STATE ${v}_{i}^{j+1}=\omega{v}_{i}^{j}+{c}_{1}{r}_{1}({\theta}_{i,best}^{j}-{\theta}_{i}^{j})+{c}_{2}{r}_{2}({G}_{best}^{j}-{\theta}_{i}^{j})$;
        \STATE ${\theta}_{i}^{j+1}={\theta}_{i}^{j}+{v}_{i}^{j+1}$;
        \ENDFOR
        \ENDFOR		
		
	\end{algorithmic}  
\end{algorithm}

\begin{equation}
    \begin{split}
    \label{Formula 10}
    {\theta}_{best}^{j}=\mathop{\arg\min}_{{\theta}_{i,best}^{u}}{\mathbb{E}}_{t \sim \mathcal{T}}[{f}_{Y}(t({X}_{adv}, {\theta}_{i,best}^{u}))] \\ i=1,2,...,I \quad u=1,2,...,j
    \end{split}
\end{equation}
where $j$ denotes the current number of iterations. ${\theta}_{i,best}^{j}$ and ${\theta}_{best}^{j}$ represent the individual optimal solution and the population optimal solution, respectively.

\textbf{Updating velocity and individual information.} Here, we update the velocity and individual information as follows:

\begin{equation}
    \label{Formula 11}
    {v}_{i}^{j+1}=\omega{v}_{i}^{j}+{c}_{1}{r}_{1}({\theta}_{i,best}^{j}-{\theta}_{i}^{j})+{c}_{2}{r}_{2}({\theta}_{best}^{j}-{\theta}_{i}^{j})
\end{equation}

\begin{equation}
    \label{Formula 12}
    {\theta}_{i}^{j+1}={\theta}_{i}^{j}+{v}_{i}^{j+1}
\end{equation}
where ${v}_{i}^{j+1}$ and ${\theta}_{i}^{j+1}$ denote the updated velocity and individual information, respectively. $\omega$ is the inertia factor, ${c}_{1}$ and ${c}_{2}$ are the learning factors of the particle, and ${r}_{1}$ and ${r}_{2}$ are random numbers between 0 and 1.

Algorithm \ref{algorithm1} presents the pseudocode of AdvCP, a proposed method for generating adversarial samples. This algorithm takes as input a clean image $X$, a target classifier $f$, a ground truth label $Y$, a maximum number of steps ${t}_{max}$, and hyperparameters ($\omega$, ${c}_{1}$, ${c}_{2}$) of PSO that are determined by the attacker. The algorithm searches for the physical parameters of the color projection $\theta(\mathcal{C}, {\mathcal{P}}_{l}, \mathcal{I})$ that generate an adversarial sample ${X}_{adv}$ that can fool the classifier $f$ by minimizing the confidence score ${f}_{Y}({X}_{adv})$ on the ground truth label $Y$. We set $\omega$, ${c}_{1}$, ${c}_{2}$ to 0.9, 1.6 and 2.0, respectively, based on the results of many experiments, to promote faster search for the global optimal solution. The algorithm outputs the physical parameters of the color projection $\theta$, which can be used to perform subsequent physical attacks. Please refer to Algorithm \ref{algorithm1} for more details.

\begin{figure}
\centering
\includegraphics[width=0.7\columnwidth]{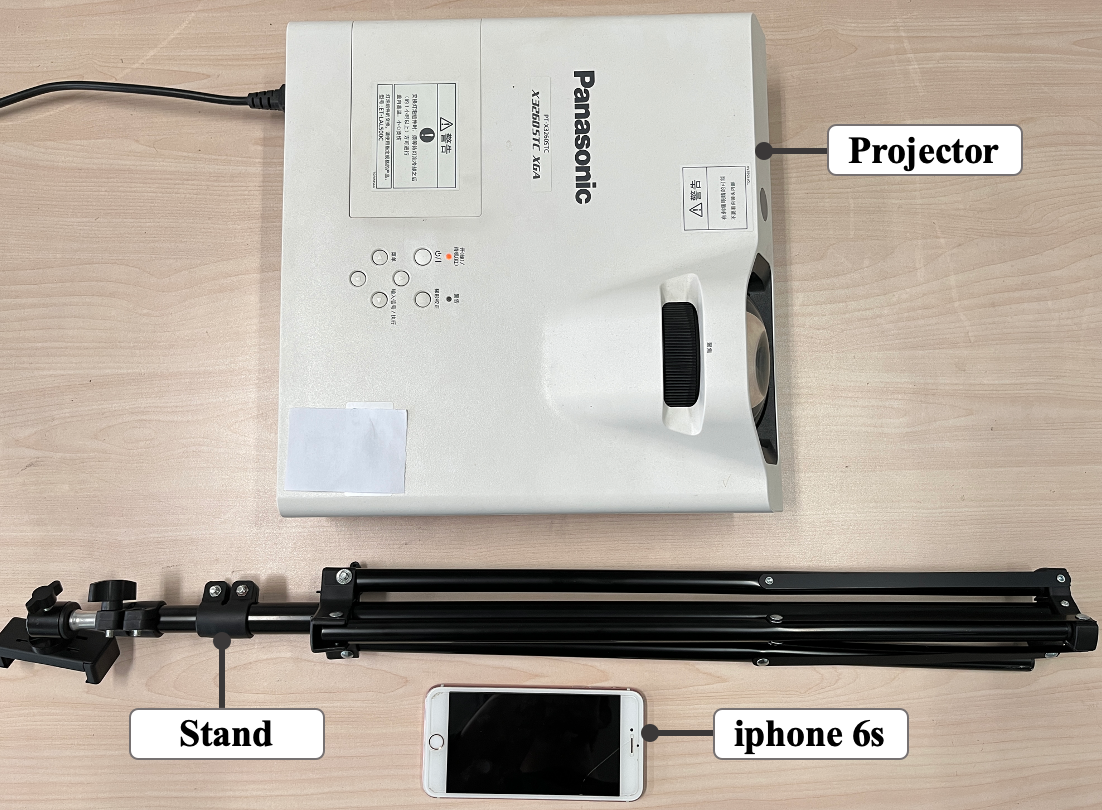} 
\caption{Experimental devices.}.
\label{figure4}
\vspace{-20pt}
\end{figure}


\section{Evaluation of AdvCP}
\label{sec4}

\subsection{Experimental settings}

We conduct experiments to test the proposed method in both digital and physical environments. Similar to AdvLB \cite{ref35}, we use ResNet50 \cite{ref40} as the target model for all experiments and randomly select 1000 correctly classified ImageNet \cite{ref47} images by ResNet50 for digital tests. For physical tests, we use a Panasonic X3260STC projector to project the color projection and an iPhone 6S to capture the photos. Our experimental devices is shown in Figure \ref{figure4}. Experiments verify that different projector and camera devices do not affect the effectiveness of the proposed AdvCP. The attack success rate (ASR) is used as a criterion to report the effectiveness of AdvCP for all experiments, which is defined as follows:

\begin{equation}
\label{eq:Positional Encoding}
\begin{split}
    &{\rm ASR}(X) = 1-\frac{1}{N}\sum_{i=1}^{N}F({y}_{i})\\
    &F({y}_{i})=
        \begin{cases}
        1 & {y}_{i} = {Y}_{i} \\
        0 & otherwise
        \end{cases}
\end{split}
\end{equation}
where $N$ is the number of clean samples that can be correctly classified by $f$ in the dataset $X$, ${Y}_{i}$ represents the ground truth label of the $i-th$ sample, ${y}_{i}$  is the label predicted under attacking.

\begin{figure*}
\centering
\includegraphics[width=1\linewidth]{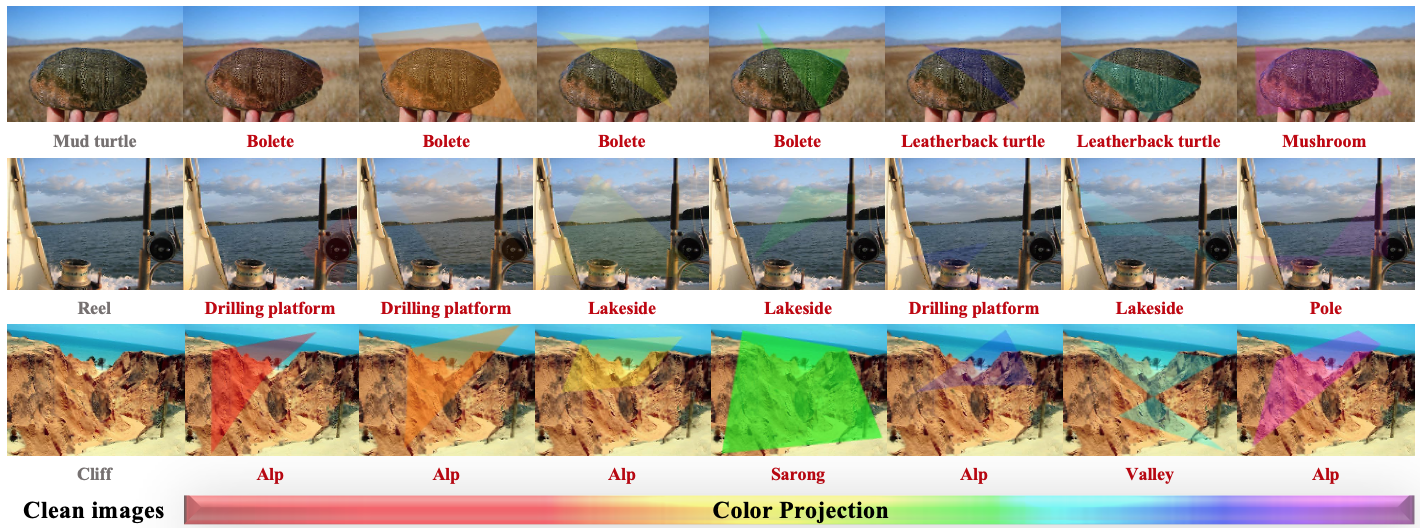} 
\caption{Adversarial samples generated by AdvCP. }.
\label{figure5}
\vspace{-10pt}
\end{figure*}

\subsection{Evaluation of effectiveness}
\textbf{Digital test.} To validate the effectiveness of AdvCP in digital environments, we conduct experiments on 1000 images that ResNet50 correctly classified. Our attack achieve an impressive success rate of 97.6\% (compared to the untargeted ASR of 95.1\% in AdvLB \cite{ref35} and the targeted ASR of 49.6\% in \cite{ref38}). We present the digital samples in Figure \ref{figure5}. The first column shows the target images, while the remaining columns display the generated adversarial samples, where various color projections, including red, orange, yellow, green, blue, indigo, and purple, are added to clean images. Our method effectively fools the classifier without altering the semantic content of the target objects. For example, adding a purple ($\mathcal{C}(255,0,255)$) color projection to the clean sample caused the mud turtle to be misclassified as a mushroom. Besides, covering the samples with various color projections led most of the cliff to be misclassified as alp. Overall, AdvCP demonstrates adversarial effectiveness in digital environments.

\begin{figure}
\centering
\includegraphics[width=1\columnwidth]{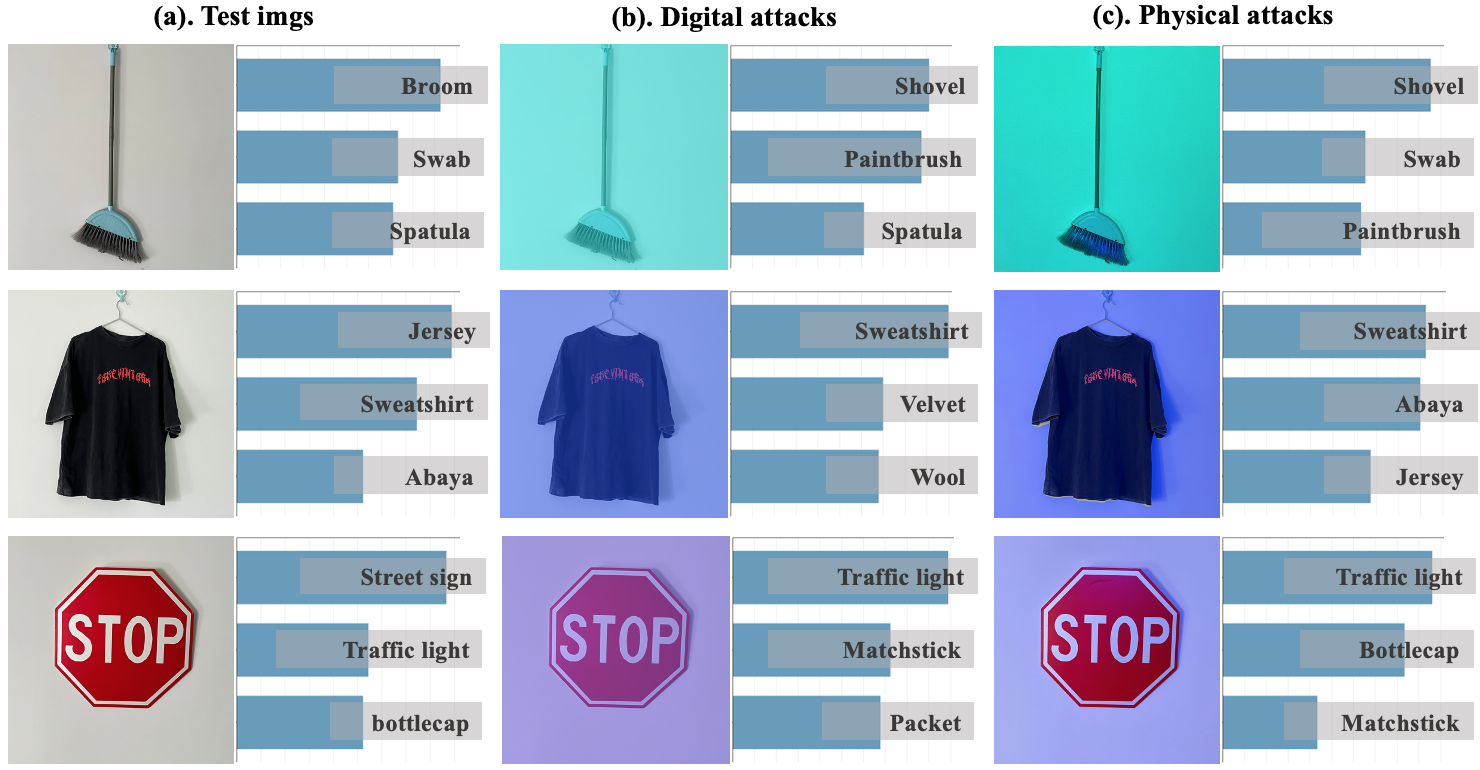} 
\caption{Indoor test.}.
\label{figure6}
\vspace{-10pt}
\end{figure}

\textbf{Physical test.} To ensure the rigor of AdvCP, we design strict experiments to evaluate its performance in physical tests. As the robustness of physical attacks can be affected by environmental noise, we conduct indoor and outdoor tests separately. The indoor test is designed to avoid the influence of outdoor noise, while the outdoor test aim to reflect the performance of AdvCP in real-world scenarios.

For the indoor test, we carefully select target objects such as ‘Broom’, ‘Jersey’, ‘Street sign’, and generated 74 adversarial samples. The results show that AdvCP achieve an ASR of 100\%, which is consistent with the performance reported in AdvLB \cite{ref35}. The experimental results are presented in Figure \ref{figure6}. Notably, the simulated color projections exhibit good consistency with the physical projections. However, there exists a slight discrepancy between the simulated and physical color projections, which can be avoided using EOT \cite{ref31}.

\begin{figure*}
\centering
\includegraphics[width=1\linewidth]{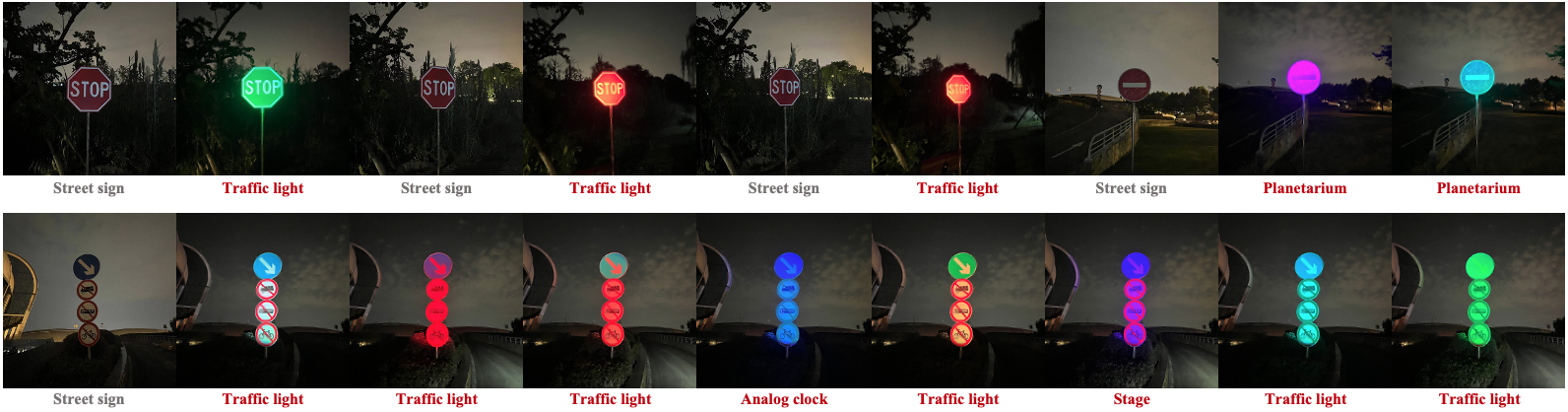} 
\caption{Outdoor test.}.
\label{figure7}
\vspace{-10pt}
\end{figure*}

In the outdoor test, we select 'Street sign' as the target object and create 112 adversarial samples, achieving an ASR of 82.14\% (compared to the ASR of 77.43\% in AdvLB \cite{ref35} and ASR of 73.26\% in \cite{ref38}). The results are presented in Figure \ref{figure7}, which displays the generated adversarial samples in the outdoor environment. The experimental results demonstrate that the proposed AdvCP method is effective in leading advanced DNNs to misclassifications by adding optimized color projection interference to the target objects. Furthermore, to simulate real-world conditions, we conduct outdoor tests on 'Stop sign' from different angles, and the results show that AdvCP can perform effective physical attacks on target objects at various angles.

We summarize the experimental results of our approach with baselines as shown in Table \ref{comparison}. It can be seen that our method is more efficient than baselines in both digital and physical environments. In conclusion, based on the comprehensive experimental results, we can confidently state that AdvCP is an effective method for adversarial attacks in both digital and physical environments. Through rigorous testing, we demonstrate the ability of AdvCP to fool advanced DNNs without altering the semantic information of the target objects. Moreover, the indoor and outdoor tests provide additional evidence of the robustness of AdvCP under different environmental conditions. Therefore, we believe that AdvCP has significant potential for use in various real-world applications, such as object recognition systems, autonomous driving, and surveillance systems, where robustness is critical.

\begin{table} 
	\centering
    \setlength{\belowcaptionskip}{10pt}
    \caption{Comparison of experimental results between AdvLB, adversarial camera sticker (AdvCS for short) and AdvCP.}.
    \label{comparison}
	\begin{tabular}{ccc|cc}

    \hline
		\multirow{2}*{Method} & \multicolumn{2}{c}{Digital} & \multicolumn{2}{c}{Physical ASR(\%)}\\
		\cline{2-5}
		~ &  ASR(\%) & Query & Indoor & Outdoor \\
		\hline
        AdvLB\cite{ref35}&95.1&834.0&\textbf{100}&77.4\\
        \hline
        AdvCS\cite{ref38}&49.6&$\varnothing$&$\varnothing$&73.3\\
        \hline
        AdvCP&\textbf{97.6}&\textbf{597.4}&\textbf{100}&\textbf{82.1}\\
        \hline

\end{tabular}
\end{table}

\begin{table}
\centering
\setlength{\belowcaptionskip}{0.5cm}
\caption{\label{Table 1} Evaluation across various classifiers.}
\begin{tabular}{cccc}
\hline

$f$ & Top-1 Accuracy(\%) & ASR(\%) & Query\\
\hline

Inception v3 & 87.6 & 85.4 & 273.4 \\
\hline

VGG19 & 91.5 & 93.9 & 165.5\\
\hline

ResNet101 & 96.1 & 93.7 & 174.8\\
\hline

GoogleNet & 85.3 & 87.2 & 253.6\\
\hline

AlexNet & 79.6 & 98.4 & 93.6\\
\hline

MobileNet & 89.7 & 89.1 &196.1\\
\hline

DenseNet & 90.8 & 94.4 &132.8\\

\hline

\end{tabular}
\end{table}

\begin{table}
    \setlength{\belowcaptionskip}{0.5cm}
    \centering
    \caption{\label{Table 2}Transferability of AdvCP.}
    \begin{tabular}{ccc}
    \hline
    $f$ & Digital (\%) & Physical (\%) \\
    \hline
    Inception v3 & 70.18 & 91.30\\
    \hline
    VGG19 & 61.07 & 95.65\\
    \hline
    ResNet101 & 52.87 & 95.65\\
    \hline
    GoogleNet & 67.11 & 100\\
    \hline
    AlexNet & 82.48 & 100\\
    \hline
    MobileNet & 64.36 & 95.65\\
    \hline
    DenseNet & 53.79 & 95.65\\
    \hline
    \end{tabular}
\end{table}

\subsection{Evaluation of stealthiness}

As previously mentioned, we opt for color projection as our physical perturbation method to create a more natural-looking physical sample, which would make it easier to overlook the perturbation by human observers. In urban environments, color beams directed towards street signs can induce a perceptual phenomenon in which people become habituated to their presence, leading to reduced sensitivity to these perturbations. However, exposure to color beams can also disrupt the functioning of DNN-based systems, potentially causing errors or system failures. Figure \ref{figure7} shows that our physical samples resemble natural color beams falling on a street sign, and human observers have difficulty distinguishing between natural and artificial color beams.
Moreover, as shown in Figure \ref{figure2}, the physical perturbations generated by AdvCP are more stealthy than the baseline. AdvCP's light-speed attack provides greater temporal stealthiness compared to RP2 \cite{ref24}, which always adheres to the target object's surface, whereas AdvCP can control the light source, generating the physical perturbation only when the attack is carried out. Compared to AdvLB \cite{ref35}, the physical samples generated by AdvCP are more natural, allowing for better spatial stealthiness in our approach. When cardboard is used for a shadow attack \cite{ref37}, it loses its spatial stealthiness when placed in front of the road sign, making human observers suspicious. In contrast, our approach places the projector far away from the target object, making AdvCP more stealthy than shadow attacks.
In general, our approach results in a more stealthy attack than the baseline, with AdvCP's physical perturbation resembling natural color beams falling on a street sign, and with greater temporal and spatial stealthiness.

\subsection{Evaluation of robustness}

\textbf{Deploy AdvCP to attack advanced DNNs}.
We conduct a thorough evaluation of the proposed AdvCP's robustness in a black-box setting against various classifiers, including advanced DNNs such as Inception v3 \cite{ref45}, VGG19 \cite{ref41}, ResNet101 \cite{ref40}, GoogleNet \cite{ref42}, AlexNet \cite{ref44}, MobileNet \cite{ref43}, and DenseNet \cite{ref39}. The dataset used for this evaluation consists of 1000 images selected from ImageNet that can be correctly classified by ResNet50. Table \ref{Table 1} presents the ASR of our method against advanced DNNs. Notably, the black-box attack test revealed that AlexNet is the most vulnerable classifier with a 98.4\% ASR and an average of 93.6 queries. The results in Table \ref{Table 1} demonstrate that AdvCP exhibits an adversarial effect on various models by more than 85\% in the black-box setting, underscoring the robustness of our proposed method.

\begin{figure*}[t]
\centering
\includegraphics[width=1\linewidth]{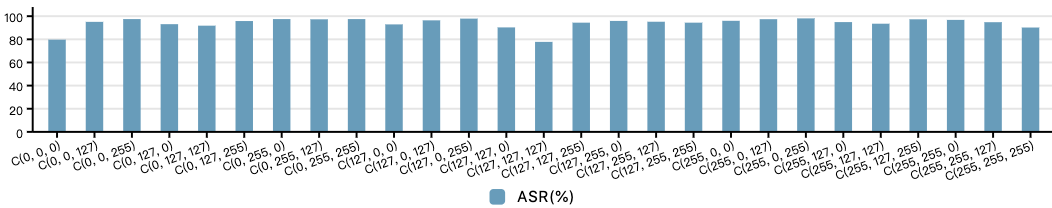} 
\caption{Ablation of $\mathcal{C}(r, g, b)$.}.
\label{figure8}
\vspace{-10pt}
\end{figure*}

\begin{table*}[h]
    \setlength{\belowcaptionskip}{0.4cm}
    \centering
    \caption{\label{Table 3}Ablation of $\mathcal{I}$.}
    \begin{tabular}{cccccccccc}
    \hline
    $\mathcal{I}$ & 0.1 & 0.2 & 0.3 & 0.4 & 0.5 & 0.6 & 0.7 & 0.8\\
    \hline
    ASR & 24.6\% & 46.1\% & 61.9\% & 74.1\% & 85.1\% & 90.4\% & 94.1\% & 97.5\% \\
    \hline
    \end{tabular}
\end{table*}

\begin{table*}[h]
    \setlength{\abovecaptionskip}{0.5cm}
    \setlength{\belowcaptionskip}{0.4cm}
    \centering
    \caption{\label{Table 4}Ablation of ${\mathcal{P}}_{l}$.}
    \begin{tabular}{cccccccc}
    \hline
    ${\mathcal{P}}_{l}$ & 3 & 4 & 5 & 6 & 7 & 8 & 9 \\
    \hline
    ASR & 79.8\% & 84.1\% & 87.4\% & 87.6\% & 88.9\% & 91.3\% & 92.1\% \\
    \hline
    \end{tabular}
\end{table*}

\textbf{Transferability of AdvCP.}
In this study, we investigate the transferability of AdvCP in both digital and physical environments against advanced DNNs \cite{ref39,ref40,ref41,ref42,ref43,ref44,ref45}. To this end, we use the adversarial samples generated by AdvCP that successfully attacked resnet50 as the dataset. The results of our experiments are presented in Table \ref{Table 2}. The data indicate that AdvCP exhibits remarkable attack transferability in the digital environment. The success rate of its transfer attacks on advanced DNNs exceeds 50\% and even achieves an impressive ASR of 82.48\% for AlexNet. In the physical test, AdvCP also demonstrates exceptional attack transferability, paralyzing almost all of the advanced DNNs. Our findings suggest that AdvCP can facilitate efficient physical attacks against advanced DNNs without requiring any knowledge of the model, by exploiting the transferability of adversarial samples.

Based on the experimental results, it can be concluded that AdvCP is capable of carrying out effective physical attacks in a black-box setting. As demonstrated in Table \ref{Table 2}, AdvCP exhibits remarkable physical adversarial transferability. This feature provides attackers with a high degree of flexibility, allowing them to perform robust physical attacks without the need for any prior knowledge of the targeted model. Given the outstanding adversarial impact of AdvCP on vision-based systems in real-world scenarios, we strongly recommend further attention and research on this method.

\section{Discussion}
\label{sec5}

\subsection{Ablation study}
We conducted a series of experiments to investigate the impact of different physical parameters on the adversarial effect of AdvCP, including intensity ($\mathcal{I}$), the number of edges (${\mathcal{P}}_{l}$), and color ($\mathcal{C}(r, g, b)$).

Intensity $\mathcal{I}$: Here, we perform experiments on a dataset of 1000 images that can be correctly classified by ResNet50. Our findings reveal that increasing the intensity $\mathcal{I}$ results in a stronger adversarial effect but also decreases stealthiness. Specifically, we study the adversarial effect of color projection with intensities ranging from 0.1 to 0.8. Table \ref{Table 3} presents the attack success rates for each intensity of color projection, highlighting the aggressive nature of AdvCP even at a low intensity.

The number of edges ${\mathcal{P}}_{l}$: From a theoretical perspective, ${\mathcal{P}}_{l}$ could be any polygon. However, increasing the number of edges can make the generated perturbation look more unnatural, thereby violating the stealthiness requirement. Therefore, we limit the attack to quadrilaterals, as shown in Table \ref{Table 4}. This approach has proven to be effective, with successful adversarial samples even generated using trilaterals.

Color $\mathcal{C}(r, g, b)$: In this study, we investigate the impact of $\mathcal{C}(r, g, b)$ on AdvCP's performance. We apply 27 different colors of color projection to perform digital attacks on ResNet50. The ASR of each color is depicted in Figure \ref{figure8}. Our results demonstrate that $\mathcal{C}(255, 0, 255)$ yields the highest ASR of 98.1\%, whereas $\mathcal{C}(127, 127, 127)$ produces the lowest ASR of 77.8\%.

\subsection{Class activation mapping}
\begin{figure}
\centering
\includegraphics[width=1\linewidth]{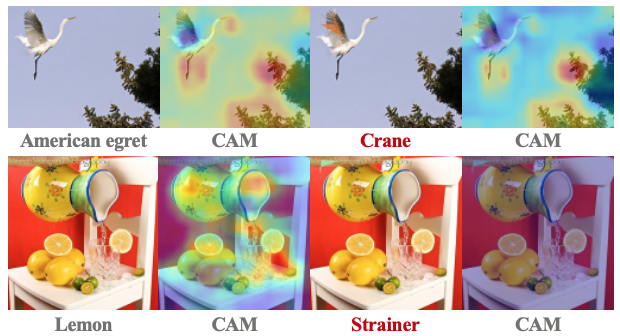} 
\caption{CAM for images.}.
\vspace{-0.5cm}
\label{figure9}
\end{figure}

Zhou et al. \cite{ref48} conducted a comprehensive investigation of the global mean pooling layer and its potential to enhance the localization capabilities of convolutional neural networks in image labeling tasks. Through extensive experimentation, they demonstrated that deep neural networks can effectively identify image regions for positioning even without specific training for this task. Their study utilized class activation mapping (CAM) to highlight the model's attention during image classification. In our work, we also employ CAM to visualize the model's attention. As depicted in Figure \ref{figure9}, by introducing optimized color projections, even in the corners of clean images, the target model exhibits a stronger bias towards categories such as 'Crane', 'Strainer', resulting in incorrect TOP-1 predictions.

\begin{figure*}[t]
\centering
\includegraphics[width=1\linewidth]{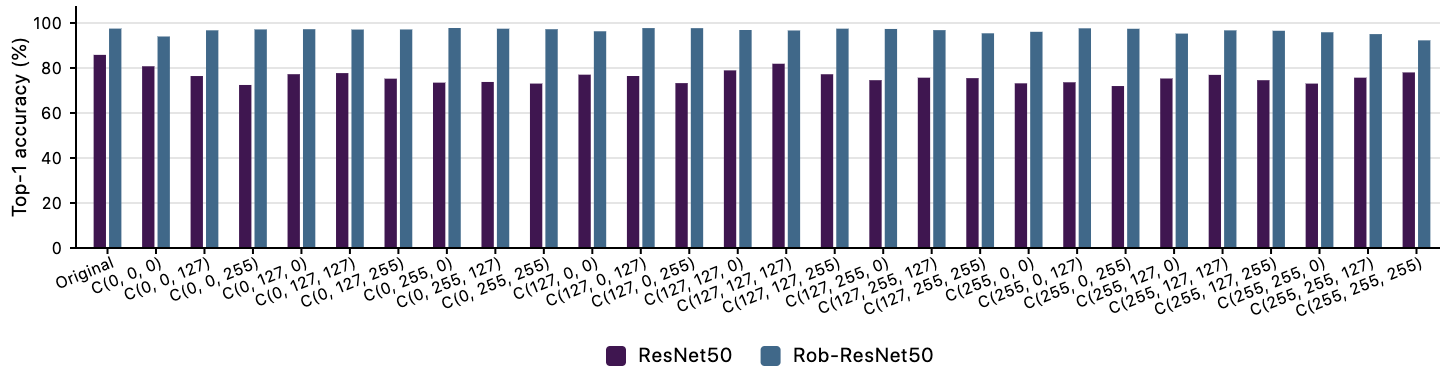} 
\caption{ResNet50 vs. Rob-ResNet50.}.
\label{figure10}
\end{figure*}


\subsection{Defense of AdvCP}

In this study, we describe our efforts to both demonstrate the potential risks posed by AdvCP and defend against this type of attack using adversarial training. To conduct a rigorous analysis of our defense strategy, we create a larger dataset, which we refer to as ImageNet-ColorProjection (ImageNet-CP).
To create this dataset, we begin by randomly selecting 50 clean samples from each of the 1000 categories in ImageNet \cite{ref47}, resulting in a total of 50,000 clean samples. We then apply 27 color projections to each clean sample with a value of $\mathcal{I}=0.7$, resulting in 1.35 million adversarial samples in the final dataset.

To conduct our adversarial training, we utilize the ImageNet-CP dataset. We employ the torchvision library to train the ResNet50 model, which we refer to as Rob-ResNet50. The training process is optimized on 3 2080Ti GPUs using ADAM with an initial learning rate of 0.01.
The experimental results can be found in Figure \ref{figure10}. As can be observed, Rob-ResNet50 achieves a classification accuracy of over 90\% for adversarial samples. We deploy AdvCP to attack Rob-ResNet50 with ${\mathcal{P}}_{l}$=4, $\mathcal{I}$=0.7, and achieve 78.2\% ASR with 329.6 average queries (ResNet50: 97.6\% ASR, 107.4 average queries). This implies that while adversarial training can reduce AdvCP's ASR and increase AdvCP's attack time cost, it cannot completely defend against AdvCP.

\section{Conclusion}
\label{sec6}
This paper introduces a new physical attack method, AdvCP, that utilizes the instantaneous nature of light to conduct black-box attacks. We evaluate the effectiveness, stealthiness, and robustness of AdvCP in our study. Rigorous experimental design and comprehensive experimental results demonstrate the effectiveness of AdvCP in both digital and physical environments. We also showcase the stealthiness of our method in terms of temporal and spatial stealthiness by comparing the generated physical sample with a baseline. To further evaluate the robustness of AdvCP, we employ AdvCP to launch attacks on advanced DNNs, and then verify the attack transferability of AdvCP. The results indicate that AdvCP poses a significant security threat to real-world scenarios, as its simple operation and superior attack transferability reflect the danger it poses to vision-based systems in low-light environments.
Our work provides insights into future physical attacks, such as using light as physical perturbations instead of stickers, which would improve the flexibility of physical attacks. The proposed AdvCP method is a valuable complementary approach to recent physical attacks, and our findings shed new light on the development of physical attacks in the future.

In the future, we plan to apply the proposed AdvCP to adapt to different tasks, including but not limited to object detection and domain segmentation. Additionally, we will expand our research to investigate other types of light-based physical attacks, such as adversarial reflected light and adversarial spot light.
Furthermore, developing effective defense strategies against light-based attacks is a promising research direction that we intend to pursue. By exploring new ways to mitigate these attacks, we hope to contribute to the development of more secure vision-based systems in the future.



{\small
\bibliographystyle{ieee_fullname}
\bibliography{egbib}
}

\end{document}